\documentclass[reprint,amsmath,amssymb,aps]{revtex4-1}

\usepackage{graphicx}

\newcommand*{\ie}{i.\,e.}

\newcommand*{\mbb}{\mathbb}

\newcommand*{\diff}{\mathop{}\!\mathrm{d}}
\newcommand*{\pd}{\partial}
\newcommand*{\im}{\ensuremath{\mathrm{i}}}
\newcommand*{\e}{\mathop{\mathrm{e}}\nolimits}
\newcommand*{\const}{\ensuremath{\mathrm{const}}}

\newcommand*{\Four}{\mathcal{F}}
\newcommand*{\Lapl}{\mathcal{L}}
\newcommand*{\ctrw}{\mathrm{RW}}
\newcommand*{\dctrw}{\mathrm{DRW}}
\newcommand*{\drwg}{\mathrm{DRWG}}
\newcommand*{\drwb}{\mathrm{DRWB}}
\newcommand*{\te}{\mathrm{TE}}
\newcommand*{\de}{\mathrm{DE}}
\newcommand*{\Gauss}{\mathrm{G}}
\newcommand*{\Bern}{\mathrm{B}}
\newcommand*{\discr}{\mathrm{d}}
\newcommand*{\singul}{\mathrm{s}}
\newcommand*{\regul}{\mathrm{r}}

\numberwithin{equation}{section}

\begin{document}

\preprint{APS/123-QED}

\title{The delayed uncoupled continuous-time random walks\\ do not provide a model for the telegraph equation}

\author{S.\,A.\,Rukolaine}%
\email{rukol@ammp.ioffe.ru}
\author{A.\,M.\,Samsonov}%
\email{samsonov@math.ioffe.ru}
\affiliation{%
  The Ioffe Physical Technical Institute of the Russian Academy of Sciences, 26~Polytekhnicheskaya, St.\,Petersburg, 194021, Russia
}%

\date{\today}

\begin{abstract}
It has been alleged in several papers that the
so called delayed continuous-time random walks (DCTRWs) provide a
model for the one-dimensional telegraph equation at microscopic
level.
This conclusion, being widespread now, is strange, since the telegraph equation describes
phenomena with finite propagation speed, while the velocity of the
motion of particles in the DCTRWs is infinite. In this paper we
investigate how accurate are the approximations to the DCTRWs
provided by the telegraph equation. We show that the diffusion
equation, being the correct limit of the DCTRWs, gives better
approximations in $L_2$ norm to the DCTRWs than the telegraph equation. We
conclude therefore that, first, the DCTRWs do not provide any
correct microscopic interpretation of the one-dimensional
telegraph equation, and second, the kinetic (exact) model of the
telegraph equation is different from the model based on the
DCTRWs.
\begin{description}
\item[PACS numbers]
05.40.-a, 05.40.Fb, 05.60.-k, 05.60.Cd, 02.30.Jr, 02.30.Mv
\end{description}
\end{abstract}

\pacs{Valid PACS appear here}

\maketitle

\section{Introduction}

The continuous-time random walks  (CTRWs)
\cite{MontrollWeiss65,ScherLax73,MontrollShlesinger84,Hughes95}
present sufficiently wide and general class of random walks,
describing random motions of particles, when both the waiting
(sojourn) time between successive jumps and the jump length
(including its direction) are generally coupled random variables.
The CTRWs have been widely applied in building up models of
anomalous diffusion and transport in physics
\cite{HavlinBenAvraham87,BouchaudGeorges90,MetzlerKlafter00,Zaslavsky02,MetzlerKlafter04,HenryEtAl06,AnomTransport08,AbadEtAl10,Fedotov11} (to name a few)
and biology \cite{MetzlerKlafter04,HornungEtAl05,YusteEtAl10}, and
in economical problems \cite{Scalas06}.

In their original form \cite{MontrollWeiss65} the CTRWs describe
random walks, or, more precisely, jumps, on lattices when only the
waiting time is a random variable. In the general case of the uncoupled CTRWs both waiting time and jump
length are independent random variables, this leads to the jump
model of the CTRWs.
The kinetic model of random
walks, often called random flights, originates from the
Pearson--Rayleigh random walks \cite{Hughes95}. In this model the
particle moves  with constant velocity for a random time along
straight lines between points where it  changes randomly the
direction of movement. The random flights can be considered both in the framework of the
generalized linear transport (kinetic) equation \cite{Alt80} and in the
framework of the CTRWs, when the waiting time and jump length are
coupled random variables. The latter is the velocity or velocity-jump
model of the CTRWs \cite{ZumofenKlafter93}.

The jump model of the CTRWs has gained high popularity due to its greater simplicity than the kinetic (velocity-jump) model. However, in the jump model the waiting time can generally be arbitrarily small or/and the jump length can be arbitrary long (Levy flights), and, therefore, a walking particle can move with infinite velocity. Thus, the jump model may, in contrast to the kinetic one, violate the principle of causality.

This is similar to the classic \emph{diffusion paradox} \cite{MoninYaglom71,JosephPreziosi89,JosephPreziosi90,MasoliverWeiss96,JouEtAl10} related to the conventional diffusion equation, which is widely used for approximate
macroscopic description of nonanomalous diffusion and Brownian motion. The difference is that in classic Einstein's model of Brownian motion \cite{Einstein56}, leading to the diffusion equation, the particle moves with finite velocity. In fact Einstein's model is nothing but the symmetric Bernoulli random walk \cite{Feller68}, which can be considered as a degenerate case of the CTRWs in the frameworks of both the jump and kinetic models.

The telegraph equation
\cite{MoninYaglom71,JosephPreziosi89,MasoliverWeiss96,Weiss02,JouEtAl10} was proposed as an alternative to the diffusion equation. In contrast to the latter, which is parabolic, the telegraph equation is
hyperbolic, providing the finite speed of signal propagation.
Two- and three-dimensional telegraph equations meet some formal
problems since Green's functions can become negative
\cite{PorraEtAl97}. Though in the case of thermal conduction this
can be fixed by imposing restrictions on the heat flux
\cite{JouEtAl10}. Yet the one-dimensional telegraph equation avoids the diffusion paradox and provides better
model for nonanomalous diffusion than the one-dimensional
diffusion equation.

In Ref.\,\cite{CamposMendez09} the authors discussed microscopic
models of the telegraph equation. Aside from the kinetic model
they proposed another model based on uncoupled CTRWs with the
waiting time distributed according to the gamma law $\left(
t/\theta^2 \right) \e^{-t/\theta}$ and the jump length having the
finite second moment. The authors called the CTRWs with the
gamma-distributed waiting time delayed CTRWs (DCTRWs). Similar
arguments were used in Ref.\,\cite{MendezCasasVazquez08} when
deriving the telegraph equation with reaction. In
Ref.\,\cite{FortMendez02} the telegraph equation with reaction was
derived formally from the CTRW with exponentially distributed
waiting time, and  later, in Ref.\,\cite{FortPujol08}, --- from
the CTRW with a general waiting time.

Earlier, the telegraph equation  was derived in
Ref.\,\cite{KenkreEtAl73} from uncoupled CTRWs with the waiting
time having more general distribution than the gamma law. We call
these CTRWs for brevity also delayed CTRWs (DCTRWs).

The conclusions that the DCTRWs simulates
the telegraph equation at the microscopic level are strange. Indeed,
the telegraph equation describes phenomena with finite propagation
speed, while the particle in the
DCTRWs may move with infinite velocity, \ie, the DCTRWs do not reflect the principal
peculiarity of the telegraph equation. In spite of this
discrepancy, one may believe that
the telegraph equation still gives a precise approximation to the DCTRWs.

In this paper we investigate how accurate are the approximations
to the DCTRWs provided by the telegraph equation. We show that the
diffusion equation, being the correct limit of the DCTRWs, gives
better approximations to the DCTRWs than the telegraph equation.
One can conclude, therefore, that the DCTRWs cannot be a
model for the telegraph equation at microscopic level.

The rest of the paper is organized as follows. 
In Section~\ref{sec:CTRW} we recall some facts concerning the uncoupled CTRWs. 
Section~\ref{sec:TeleApprox} shows how the
telegraph equation was derived from the DCTRWs in
Ref.\,\cite{CamposMendez09,MendezCasasVazquez08,FortMendez02,FortPujol08,KenkreEtAl73}.
The asymptotic behaviour of the DCTRWs is discussed in Section~\ref{sec:AsymptBehav}. 
In Section~\ref{sec:NumerCompar} we compare the telegraph and diffusion
approximations to two (discrete- and continuous-space) DCTRWs, completely described in Section~\ref{sec:TwoParticul}, numerically, while asymptotic comparison of the approximations to the continuous-space DCTRW at long times is performed in Section~\ref{sec:AsymptCompar}.
Section~\ref{sec:Concl} contains some conclusive remarks.

\section{Uncoupled  continuous-time random walks}
\label{sec:CTRW}

In this section we briefly formulate some facts concerning the
CTRWs for further reference.

We adopt the one-dimensional model and consider
uncoupled CTRWs, when the waiting time and jump length are
independent random variables, \ie, the joint probability density
$\varphi(x,t)$ for a particle to jump a distance $x$ after waiting
a time $t$ is given by
\begin{equation*}
  \varphi(x,t) = \psi(t) \lambda(x),
  \quad x \in \mathbb{R}, \quad t \geq 0,
\end{equation*}
where $\psi(t)$ and $\lambda(x)$ are probability densities for
the waiting time and jump length, respectively, with obvious
normalizations $\int_0^\infty \psi(t) \diff{t} = 1$ and
$\int_{-\infty}^\infty \lambda(x) \diff{x} = 1$.

We consider here CTRWs in a continuous space,
however, they can be similarly described on a
lattice, as it was originally proposed in
\cite{MontrollWeiss65}, with obvious modifications.

The CTRW is described by the density $\rho_\ctrw^{}(x,t)$, which
is a probability density that the particle is at the point $x$ at
the time $t$, so that $\int_{-\infty}^\infty \rho_\ctrw^{}(x,t)
\diff{x} = 1$. The density $\rho_\ctrw^{}(x,t)$ is given
by the integral equation
\cite{KlafterEtAl87,HenryEtAl06}
\begin{multline}
  \label{eq:CTRWIEq}
  \rho_\ctrw^{}(x,t) =
  \varPsi(t) \delta(x)\\[1ex]
  \shoveright{+ \int_0^t \int_{-\infty}^\infty \varphi(x-x',t-t') \rho_\ctrw^{}(x',t') \diff{x'} \diff{t'}}\\[1ex]
  \shoveleft{= \varPsi(t) \delta(x)}\\[1ex]
  + \int_0^t \psi(t-t') \left[ \int_{-\infty}^\infty \lambda(x-x') \rho_\ctrw^{}(x',t') \diff{x'} \right] \diff{t'},
\end{multline}
where $\delta(\cdot)$ is the Dirac delta function, and
\begin{equation*}
  \varPsi(t) =
  1 - \int_0^t \psi(t') \diff{t} \equiv
  \int_t^\infty \psi(t') \diff{t},
  \quad t \geq 0,
\end{equation*}
is the survival probability, \ie, the probability that a particle
stays at the same position for the time $t$. Here one assumes that
at the initial moment $t=0$ the particle was at the point $x=0$,
\ie, $\rho_\ctrw^{}|_{t=0} = \delta(x)$.

The Fourier--Laplace transform of the equation \eqref{eq:CTRWIEq},
denoted  by $\Four{\Lapl\cdot}$, implies the Montroll--Weiss
formula \cite{MontrollWeiss65,MontrollShlesinger84}
\begin{multline}
  \label{eq:MWEq}
  \Four{\Lapl\rho}_\ctrw^{}(k,s) =
  \frac{\Lapl\varPsi(s)}{1 - \Four{\Lapl\varphi}(k,s)} =
  \frac{1-\Lapl\psi(s)}{s \,[1 - \Four{\Lapl\varphi}(k,s)]}\\[1ex]
  = \frac{1-\Lapl\psi(s)}{s \,[1 - \Four{\lambda}(k) \Lapl\psi(s)]},
\end{multline}
where
\begin{equation*}
  \Four{\lambda}(k) =
  \int_{-\infty}^\infty \lambda(x) \e^{\im k x} \diff{x}
\end{equation*}
is the Fourier transform of $\lambda$,
\begin{equation*}
  \Lapl{\psi}(s) =
  \int_0^\infty \psi(t) \e^{-s t} \diff{t}
\end{equation*}
is the Laplace transform of $\psi$.

In the uncoupled CTRWs the density  $\rho_\ctrw^{}$ can be
expressed explicitly (such a CTRW is the random walk subordinated
to a renewal process) \cite{Feller71,ScalasEtAl04}:
\begin{multline}
  \label{eq:DensCTRW}
  \rho_\ctrw^{}(x,t) =
  \sum_{n=0}^\infty p_n(t) \lambda_n(x)\\
  \equiv
  \varPsi(t) \delta(x) + \sum_{n=1}^\infty p_n(t) \lambda_n(x)
\end{multline}
($\rho_\ctrw^{}(\cdot,t)$, with fixed $t$, is actually  a
probability density of a random sum), where $p_n(t)$ is the
probability of $n$ jumps occurring up to the time $t$, and
$\lambda_n(x)$ is the probability density of a distance $x$ from
the initial position reached by a particle after $n$ jumps. The
probability density $\lambda_n(x)$ is given by
\begin{equation*}
  \lambda_n(x) =
  \lambda^{n{*}}(x),
\end{equation*}
where $*$ means the convolution, $\lambda^{n{*}}$ means  the
$n$-fold convolution of $\lambda$ with itself, \ie,
$\lambda^{0{*}}(x) = \delta(x)$, $\lambda^{1{*}}(x) = \lambda(x)$,
and $\lambda^{n{*}}(x) = (\lambda^{(n-1){*}} * \lambda)(x)$.
Equivalently,
\begin{equation*}
  \Four{\lambda_n}(k) = [\Four\lambda(k)]^n.
\end{equation*}
The probability $p_n(t)$ is given by
\begin{equation*}
  p_n(t) =
  \left( \varPsi * \psi^{n{*}} \right)(t)
\end{equation*}
(clearly $\psi(t)=0$ and $\Psi(t)=0$ for $t<0$), in particular,
$p_0(t) = \varPsi(t)$. Equivalently,
\begin{equation*}
  \Lapl{p}_n(s) = \Lapl\varPsi(s) [\Lapl\psi(s)]^n.
\end{equation*}

This completes the set of auxiliary propositions necessary for further considerations.

\section{Telegraph approximations to the delayed continuous-time random walks}
\label{sec:TeleApprox}

Consider a family of probability densities for waiting time $t$
\cite{KenkreEtAl73,OthmerEtAl88,CamposMendez09}
\begin{multline}
  \label{eq:SPsiA}
  \psi_a(t) \\[1ex]=
  \begin{cases}
    \dfrac{a}{\theta \sqrt{1-a}} \e^{-t/\theta} \sinh \left( \dfrac{t}{\theta} \sqrt{1-a} \right),
    & 0<a<1,\\[3.5ex]
    \dfrac{t}{\theta^2} \e^{-t/\theta}, & a=1,
  \end{cases}
\end{multline}
(obviously, $\psi_1(t) = \lim_{a \to 1} \psi_a(t)$).
The Laplace transform of the density $\psi_a$ is
\begin{multline*}
  \Lapl\psi_a(s) =
  \frac{a}{(\theta s + 1)^2 - (1-a)}\\[1ex]
  \equiv
  \frac{a}{\theta^2} \frac{1}{(s + 1/\theta)^2 - (1-a)/\theta^2}.
\end{multline*}
The density $\psi_1$ was used in
Refs.\,\cite{OthmerEtAl88,CamposMendez09} (see also Ref.\,\cite{BurchLehoucq11}),  and it belongs to the
family of the gamma distributions. The density $\psi_1$ is a
probability density of a sum of two independent exponentially
distributed random variables with the probability density
$(1/\theta)\e^{-t/\theta}$, and CTRWs with such a distribution of
waiting times were called in Ref.\,\cite{CamposMendez09}
the delayed CTRWs (DCTRWs). Note that the  mean waiting time for the exponential
distribution is $\theta$, and it is $2\theta$ for the distribution
$\psi_1$. The mean waiting time for the distribution $\psi_a$ is
$2\theta/a$, \ie, the mean waiting time tends to infinity as $a
\to 0$. Completely understanding the conditional character of this
notion we call for brevity the CTRWs with the probability
densities $\psi_a$ of the waiting time also the
DCTRWs. We denote densities describing the DCTRWs by $\rho_\dctrw^{}$.

The Montroll--Weiss formula \eqref{eq:MWEq} with $\Lapl\psi_a$ implies
\begin{equation*}
  \Four{\Lapl\rho}_\dctrw^{}(k,s) =
  \frac{\theta s + 2}{s (\theta s + 2) - a [\Four\lambda(k) - 1] /\theta}.
\end{equation*}
The straightforward calculations show that the density
$\rho_\dctrw^{}$ is a solution to the equation
\begin{multline}
  \label{eq:DCTRWEq}
  \frac{\theta}{2} \frac{\pd^2\rho_\dctrw^{}(x,t)}{\pd{t}^2} + \frac{\pd\rho_\dctrw^{}(x,t)}{\pd{t}}\\[1ex]
  = \frac{a}{2 \theta} \left[ \int_{-\infty}^\infty \lambda(x-x') \rho_\dctrw^{}(x',t) \diff{x'} - \rho_\dctrw^{}(x,t) \right]
\end{multline}
with the initial value conditions
\begin{equation*}
  \left. \rho_\dctrw^{} \right|_{t=0} = \delta(x), \quad \left. \frac{\pd\rho_\dctrw^{}}{\pd{t}} \right|_{t=0} = 0.
\end{equation*}

Consider a random walk on a regular one-dimensional lattice, where
the distance between nearest-neighbour points is $\sigma$, and the
coordinates of the lattice points are $x_l = l \sigma$, $l \in
\mathbb{Z}$. Suppose that jumps are made to the nearest-neighbour
points with equal probability. This discrete-space random walk is
described by the probability density $\lambda_\discr(x) =
[\delta(x-\sigma) + \delta(x+\sigma)]/2$, and the equation
\eqref{eq:DCTRWEq} becomes
\begin{multline}
  \label{eq:DCTRWEqD}
  \frac{\theta}{2} \frac{\pd^2\rho_\dctrw^{}(x_l,t)}{\pd{t}^2} + \frac{\pd\rho_\dctrw^{}(x_l,t)}{\pd{t}}\\[1ex]
  \shoveleft{= \frac{a}{2 \theta} \bigg[ \frac{\rho_\dctrw^{}(x_l - \sigma,t) + \rho_\dctrw^{}(x_l + \sigma,t)}{2}}\\
  - \rho_\dctrw^{}(x,t) \bigg]
\end{multline}
Assuming that the density $\rho_\dctrw^{}$ is
smooth with respect to $x$,  one can expand it in the Taylor
series in $x$. If $\sigma$ is small enough then we neglect the
derivatives of an order higher than two, and
obtain the approximation for the density $\rho_\dctrw^{}$ given
by the solution of the telegraph equation (TE)
\begin{equation}
  \label{eq:TeleEq}
  \frac{\theta}{2} \frac{\pd^2\rho_\te^{}}{\pd{t}^2} + \frac{\pd\rho_\te^{}}{\pd{t}} - \frac{a \sigma^2}{4\theta} \frac{\pd^2\rho_\te^{}}{\pd{x}^2} = 0
\end{equation}
with the initial conditions
\begin{equation}
  \label{eq:TeleEqIniCond}
  \left. \rho_\te^{} \right|_{t=0} = \delta(x), \quad \left. \frac{\pd\rho_\te^{}}{\pd{t}} \right|_{t=0} = 0.
\end{equation}
The same result is obtained if jumps are made not only to the
nearest-neighbour points, the only decisive condition is
that the variance of the distribution $\lambda_\discr$ is to be
equal to $\sigma^2$.

Similar derivation of the telegraph equation from the
discrete-space random walk with the probability density $\psi_a$
for waiting time was performed in Ref.\,\cite{KenkreEtAl73}, where
the authors wrote about ``the continuum limit'' as $\sigma \to 0$.

Note that both the equations \eqref{eq:DCTRWEq} and
\eqref{eq:DCTRWEqD} describe particles, moving with infinite
velocity, while the telegraph equation describes phenomena with
finite propagation speed.

Another derivation of the telegraph approximation  is given in
Ref.\,\cite{CamposMendez09}. This is similar to the above one, but
it is performed in the Fourier--Laplace space. The derivation,
given in Ref.\,\cite{CamposMendez09} for $a=1$, is in fact as
follows. If the distribution of the jump length is symmetric and
has the finite second moment with the variance equal to
$\sigma^2$, then the Fourier transform of the probability density
of the jump length is approximately $\Four{\lambda}(k) \approx 1 -
\sigma^2 k^2 /2$ for $|\sigma k| \ll 1$. After substitution of
this $\Four{\lambda}(k)$ and $\Lapl\psi_a^{}(s)$ into the
Montroll--Weiss formula \eqref{eq:MWEq} one obtains that the
Fourier--Laplace transform of the density $\rho_\dctrw^{}$ is
approximately equal to
\begin{equation*}
  \Four{\Lapl\rho}_\dctrw^{}(k,s) \approx
  \Four{\Lapl\rho}_\te^{}(k,s) \quad\text{for}\enskip |\sigma k| \ll 1
\end{equation*}
with
\begin{equation}
  \label{eq:DensTeleEqFourLapl}
  \Four{\Lapl\rho}_\te^{}(k,s) =
  \frac{\theta s + 2}{s (\theta s + 2) + a \sigma^2 k^2 /(2\theta)}.
\end{equation}
The inverse Fourier--Laplace transform of
$\Four{\Lapl\rho}_\te^{}$ implies  that the density $\rho_\te^{}$
is a solution to the telegraph equation \eqref{eq:TeleEq} with the
initial conditions \eqref{eq:TeleEqIniCond}. The authors of the
paper \cite{CamposMendez09} call the telegraph equation
\eqref{eq:TeleEq} ``the limit of small jumps (diffusive
approximation)'' of the CTRW with the probability density $\psi_1$
\eqref{eq:SPsiA} for waiting time.

The conclusion that the solution to the problem \eqref{eq:TeleEq},
\eqref{eq:TeleEqIniCond} approximates the DCTRW is strange, since
the telegraph equation describes phenomena with finite propagation
speed. At the same time, the velocity of the motion of the
particle in the DCTRW is infinite, since waiting time can be arbitrarily small.
Moreover, the telegraph equation obviously is not ``the limit of
small jumps'' of the DCTRWs.

\section{The asymptotic behaviour of the delayed continuous-time random walks}
\label{sec:AsymptBehav}

The asymptotic behaviour of various CTRWs was studied in
Refs.\,\cite{Kotulski95,ScalasEtAl04,GorenfloMainardi05}.
For convenience we briefly repeat the
asymptotic analysis for the DCTRWs on the basis of
Refs.\,\cite{ScalasEtAl04,GorenfloMainardi05}.
To find the asymptotic behaviour of the DCTRWs the densities
$\lambda(x)$ and $\psi_a^{}(t)$ are replaced by the scaled
densities $\lambda_h(x) = \lambda(x/h)/h$, $h>0$, and
$\psi_{a,\tau}(t) = \psi_a^{}(t/\tau)/\tau$, $\tau>0$,
respectively
\cite{ScalasEtAl04,GorenfloMainardi05}. The
parameters $h$ and $\tau$ can be considered as the characteristic
step length and waiting time, respectively. The Fourier transform
of $\lambda_h$ is $\Four{\lambda_h}(k) = \Four\lambda(h k)$, and
the Laplace transform of $\psi_{a,\tau}$ is
$\Lapl{\psi_{a,\tau}}(s) = \Lapl\psi_a^{}(\tau s)$. We suppose
that the distribution of the jump length is symmetric,
$\lambda(-x) = \lambda(x)$, and has the finite second moment with
the variance equal to $\sigma^2$, then the asymptotic behaviour
for $\Four{\lambda_h}$ is $1 - \Four{\lambda_h}(k) \sim h^2
\sigma^2 k^2 /2$ as $h \to 0$, $k \in \mbb{R}$
\cite{GorenfloMainardi05}. For the density
$\psi_{a,\tau}$ one has the asymptotics $1 -
\Lapl{\psi_{a,\tau}}(s) \sim 2 \tau \theta s /a$ as $\tau \to 0$.
The latter asymptotics are also valid in general case for
$\psi_\tau(t) = \psi(t/\tau)/\tau$, if the density $\psi$ has the
finite expectation equal to $2\theta/a$
\cite{GorenfloMainardi05}. After substitution
of the above asymptotics into the Montroll--Weiss formula
\eqref{eq:MWEq}, one can see that the only possible nontrivial
limit is
\begin{multline}
  \label{eq:DensDCTRWApproxDiffEq}
  \Four{\Lapl\rho}_{\dctrw,h,\tau}^{}(k,s) \equiv
  \frac{1-\Lapl{\psi_{a,\tau}}(s)}{s \,[1 - \Four{\lambda_h}(k) \Lapl{\psi_{a,\tau}}(s)]}\\[1ex]
  \to
  \Four{\Lapl\rho}_\de^{}(k,s) \quad\text{as}\enskip h \to 0 \enskip\text{and} \enskip \tau \to 0
\end{multline}
under the scaling relation $h^2 / \tau = 1$, where
\begin{equation*}
  \Four{\Lapl\rho}_\de^{}(k,s) =
  \frac{1}{s + a \sigma^2 k^2 /(4\theta)}
\end{equation*}
The inverse Fourier--Laplace transform implies that the density
$\rho_\de^{}$ is a solution of the diffusion equation (DE)
\begin{equation}
  \label{eq:DiffEq}
  \frac{\pd\rho_\de^{}}{\pd{t}} - \frac{a \sigma^2}{4\theta} \frac{\pd^2\rho_\de^{}}{\pd{x}^2} = 0
\end{equation}
with the initial condition
\begin{equation}
  \label{eq:DiffEqIniCond}
  \left. \rho_\de^{} \right|_{t=0} = \delta(x).
\end{equation}

Thus, the asymptotic behaviour of the DCTRWs is given by the
diffusion equation \eqref{eq:DiffEq}. Note that the diffusion
equation is obtained from the telegraph equation \eqref{eq:TeleEq}
by omitting the second time derivative.

Note also that $\Four{\Lapl\rho}_\te^{}(0,s) =1/s$ and
$\Four{\Lapl\rho}_\de^{}(0,s) =1/s$, or, equivalently,
$\Four\rho_\te^{}(0,t) =1$ and $\Four\rho_\de^{}(0,t) =1$ for $t
\geq 0$, \ie, the law of conservation of particles is valid
for both the telegraph and diffusion approximations.

The other possible limits are
$\Four{\Lapl\rho}_{\dctrw,h,\tau}^{}(k,s) \to 1/s$ as $h \to 0$,
and $\Four{\Lapl\rho}_{\dctrw,h,\tau}^{}(k,s) \to 1/s$ as $\tau
\to 0$ and $h / \sqrt\tau \to 0$. The former is the limit of
infinitesimal jumps with finite mean waiting time. This is the
correct limit [to be] obtained in
Refs.\,\cite{KenkreEtAl73,CamposMendez09}. These asymptotics are
trivial, since the inverse Fourier--Laplace transform of $1/s$
gives $\delta(x)$, $t \geq 0$, which means that the particle does
not leave an initial position.

The diffusion limit, under the scaling relation $h^2 / \tau =
\const$, is valid for a variety of symmetric (unbiased) random
walks, both continuous-time and discrete-time ones
\cite{Feller68,Feller71,Hughes95,Kotulski95}.
In particular, it is valid for the
symmetric Bernoulli random walk, which is discrete both
in time and space, when the waiting time is exactly $\tau$ and the
jumps are exactly $\pm h$ with equal probabilities, \ie, the
corresponding probability densities are
\begin{multline}
  \label{eq:BernRW}
  \psi_\discr(t) = \delta(t-\tau), \quad \tau>0,
  \quad\text{and}\\[1ex]
  \lambda_\discr(x) = \frac{1}{2} [\delta(x-h) + \delta(x+h)], \quad h>0.
\end{multline}
However, in contrast to the CTRWs, in which waiting time can be
arbitrarily small and, hence, the particle moves with an infinite
speed, the speed of the motion of the particle in the Bernoulli
random walk is finite. Besides, it turns out that the solution of
the telegraph equation approximates the density for the Bernoulli
random walk better than that of the diffusion equation
\cite{Keller04}. It is necessary to note here that the
explanation, given in Ref.\,\cite{Keller04}, does not allow us to
judge whether the telegraph equation approximates the random walk
better than the diffusion equation in a wide time interval? To
check this conclusion we have performed calculations, in which the
binomial distribution, being the distribution of the walking
particle, was evaluated through the first two terms of Stirling's
asymptotic series for the gamma function
\cite{Lebedev65} $\Gamma(z) = \sqrt{2\pi/z}
\,(z/\e)^z \,[1 + 1/(12z) + O(|z|^{-2})]$ as $|z| \to \infty$, $z
\in \mbb{C}$, $|\arg z| \leq \pi-\Delta$, $\Delta>0$, which is
more accurate than conventional  Stirling's formula. These
calculations have confirmed the conclusion up to time $t
\approx 100\tau$, which seems to be asymptotic. Obviously, the
solutions of both the diffusion equation and the telegraph one do
not approximate the binomial distribution at short time.

The reason, by which the solution of the telegraph equation
approximates the Bernoulli random walk better than that of the
diffusion equation, seems to be clear: the Bernoulli particle
moves with finite velocity. However, the probability density for
the waiting time $\psi_\discr$, Eq.\,\eqref{eq:BernRW}, can be
weakly approximated with arbitrary accuracy by the gamma
distribution:
\begin{multline}
  \label{eq:PsiMu}
  \psi_\mu(t) \equiv \frac{1}{\Gamma(\mu)} \frac{\mu}{\tau} \left( \frac{\mu t}{\tau} \right)^{\mu-1} \e^{-\mu t/\tau}\\[1ex]
  \xrightarrow{\text{weakly}} \delta(t - \tau) \quad \text{as} \enskip \mu \to +\infty,
\end{multline}
where $\Gamma$ is the Gamma function; this can easily be derived
with the Laplace transform. In the random walk with the
distribution $\psi_\mu$ of the waiting time the particle moves
with infinite velocity, since the waiting time can generally be
arbitrarily small. However, for sufficiently large $\mu$ the
telegraph approximation to this random walk is
clearly better than the diffusion one, since $\psi_\mu(t)$ is very
``close'' to $\delta(t - \tau)$.

Thus, in spite of the diffusion asymptotic behaviour for the
DCTRWs and the infinite speed of the motion of the particle, the
question still remains: which of the telegraph or the diffusion
equations gives better approximation to the DCTRWs?

\section{Comparison of the telegraph and diffusion approximations to the DCTRWs}
\label{sec:NumerCompar}

To compare the telegraph and diffusion equations with the DCTRWs
we consider two particular DCTRWs. One of them is a
discrete-space random walk, the other is the continuous-space one.

The \emph{discrete-space} random walk takes place on a regular
one-dimensional lattice, where the distance between
nearest-neighbour points is $\sigma$, and the coordinates of the
lattice points are $x_l = l \sigma$, $l \in \mathbb{Z}$. Jumps are
made to the nearest-neighbour points with equal probability. This
random walk is described by the probability density
\begin{equation}
  \label{eq:LambdaB}
  \lambda_\Bern(x) = \frac{1}{2} [\delta(x-\sigma) + \delta(x+\sigma)].
\end{equation}
The subscript ${}_\Bern$ means Bernoulli (since this is the same distribution as in the Bernoulli random walk).
The variance of this distribution is equal to $\sigma^2$.

The \emph{continuous-space} random walk has the Gaussian
distribution of the jump length with the variance equal also to
$\sigma^2$:
\begin{equation}
  \label{eq:LambdaG}
  \lambda_\Gauss^{}(x) = \frac{1}{\sqrt{2 \pi} \,\sigma} \e^{- x^2 / (2 \sigma^2)},
\end{equation}
its Fourier transform is $\Four\lambda_\Gauss^{}(k) =
\e^{-\sigma^2 k^2 /2}$.  The Gaussian distribution of the jump
length was used in Ref.\,\cite{MendezCasasVazquez08}.

The densities of the discrete- and
continuous-space DCTRWs are given in Appendix~\ref{sec:TwoParticul} by Eqs.\,\eqref{eq:DensDCTRWB} and
\eqref{eq:DensDCTRWG}, respectively. The telegraph and
diffusion approximations are given by Eqs.\,\eqref{eq:DensTeleEq}
and \eqref{eq:DensDiffEq}, respectively.

All numerical results are obtained with the parameters $\sigma=1$ and $\theta=1$.

\begin{figure}[!htb]
  \centering
  \includegraphics{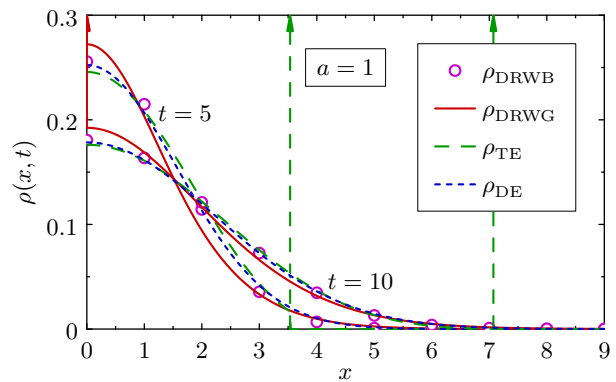}
  \caption{(Color online) The densities $\rho_\drwg^{}(x,t)$ and $\rho_\drwb^{}(x_l,t)$ and the telegraph and diffusion approximations $\rho_\te^{}(x,t)$ and $\rho_\de^{}(x,t)$ at $t=5$ and $t=10$ with $\sigma=1$, $a=1$ and $\theta=1$.
The singular terms $\rho_\drwg^\singul$ and $\rho_\te^\singul$ are depicted by vertical arrows.}
  \label{fig:Dens}
\end{figure}

\begin{figure}[!htb]
  \centering
  \includegraphics{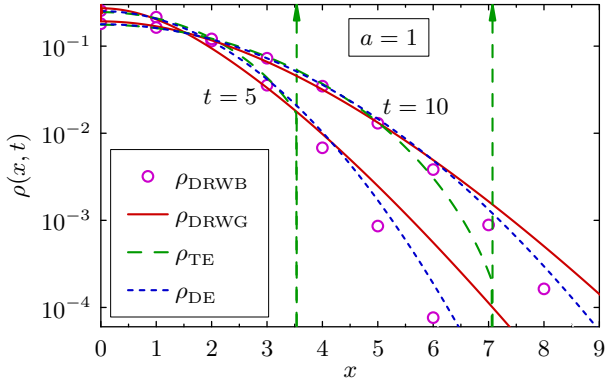}
  \caption{(Color online) The same as in Fig.\,\ref{fig:Dens}, with logarithmic scale on the vertical axis.}
  \label{fig:DensLog}
\end{figure}

\begin{figure}[!htb]
  \centering
  \includegraphics{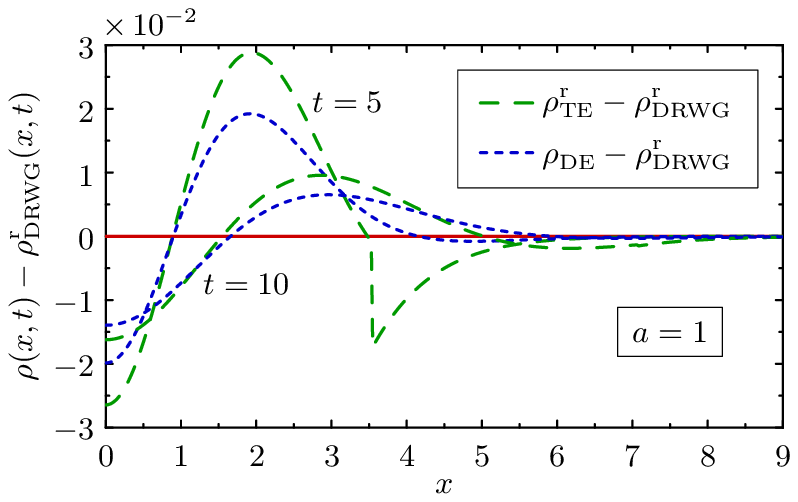}
  \caption{(Color online) The differences $\rho_\te^\regul(x,t) - \rho_\drwg^\regul(x,t)$ and $\rho_\de^{}(x,t) - \rho_\drwg^\regul(x,t)$ at $t=5$ and $t=10$ with $\sigma=1$, $a=1$ and $\theta=1$. Note that the vertical scale unit is $10^{-2}$.}
  \label{fig:DensDiscrepGauss}
\end{figure}

\begin{figure}[!htb]
  \centering
  \includegraphics{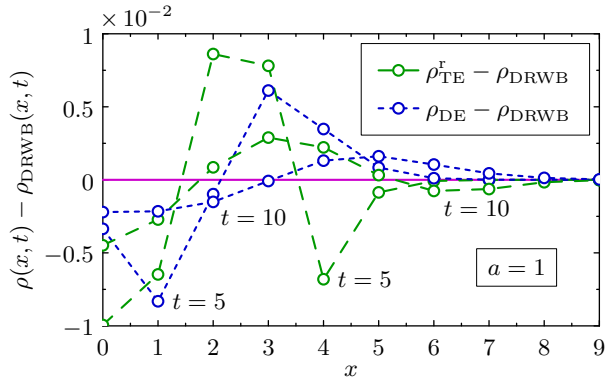}
  \caption{(Color online) The differences $\rho_\te^\regul(x_l,t) - \rho_\drwb^{}(x_l,t)$ and $\rho_\de^{}(x_l,t) - \rho_\drwb^{}(x_l,t)$ at $t=5$ and $t=10$ with $\sigma=1$, $a=1$ and $\theta=1$. Note that the vertical scale unit is $10^{-2}$.}
  \label{fig:DensDiscrepBin}
\end{figure}

Figs.\,\ref{fig:Dens} and \ref{fig:DensLog} show (in Cartesian and
logarithmic scales) the densities of the continuous- and
discrete-space DCTRWs and their telegraph and diffusion
approximations at intermediate values of time $t=5$ and $t=10$ with  $a=1$. 
Figs.\,\ref{fig:DensDiscrepGauss} and
\ref{fig:DensDiscrepBin} show differences between each of the two
approximations and the densities of the continuous- and
discrete-space DCTRWs, respectively, obtained for the same values
of time and the parameters. Calculations show that in
all the cases the $L_2$-discrepancy for the diffusion
approximation is less than that for the telegraph one. In the
case of the discrete-space random walk we calculated the
$l_2$-discrepancy. Note also that the maximum absolute value of
the difference for the diffusion approximation is in all the cases
less than that for the telegraph approximation. Besides, it is
important to emphasize that the telegraph approximation is
incorrect with respect to the velocity of the motion of the
particle, see Figs.\,\ref{fig:Dens} and \ref{fig:DensLog} for
$t=5$. The telegraph model gives the finite velocity, while it
is infinite in the DCTRWs. The diffusion model is correct in this
respect, however the diffusion paradox remains.

To assess the approximations at long times we have performed in Appendix~\ref{sec:AsymptCompar} asymptotic comparison of the telegraph and diffusion approximations to the continuous-space DCTRW through their Fourier transforms.

\begin{figure}[!htb]
  \centering
  \includegraphics{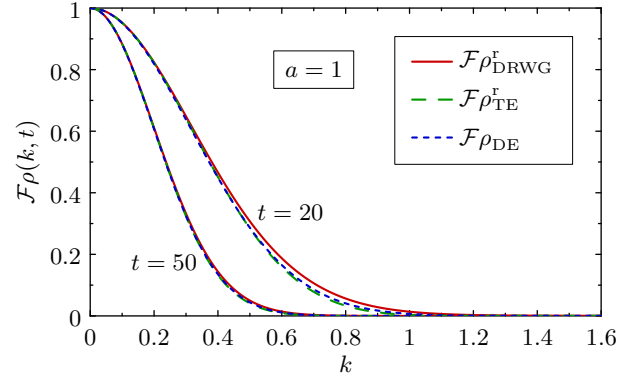}
  \caption{(Color online) Graphs resulting from the Fourier transforms $\Four\rho_\drwg^\regul(k,t)$, $\Four\rho_\te^\regul(k,t)$ and $\Four\rho_\de^{}(k,t)$ at $t=20$ and $t=50$ with $\sigma=1$, $a=1$ and $\theta=1$, \ie, $\varepsilon=0.05$ and $\varepsilon=0.02$.}
  \label{fig:FourHarmLongT}
\end{figure}

Fig.\,\ref{fig:FourHarmLongT} shows graphs resulting from the
Fourier transforms $\Four\rho_\drwg^{}(k,t)$,
$\Four\rho_\te^{}(k,t)$ and $\Four\rho_\de^{}(k,t)$ at $t=20$ and
$t=50$ with $a=1$. These values
correspond to the values of the small parameter $\varepsilon =
0.05$ and $\varepsilon = 0.02$, respectively (see Appendix~\ref{sec:AsymptCompar}). 
Fig.\,\ref{fig:FourHarmLongTDiscrep2} shows the integrands of the
integrals \eqref{eq:Integrals} for the same values of time and
parameters $\sigma$, $a$ and $\theta$ as in
Fig.\,\ref{fig:FourHarmLongT}. Note that the asymptotics
\eqref{eq:DTeleDCTRWGRegulFourZetaAsy} and
\eqref{eq:DDiffDCTRWGRegulFourZetaAsy} provide very good
approximations (not shown in the figure) to the integrands.

\begin{figure}[!htb]
  \centering
  \includegraphics{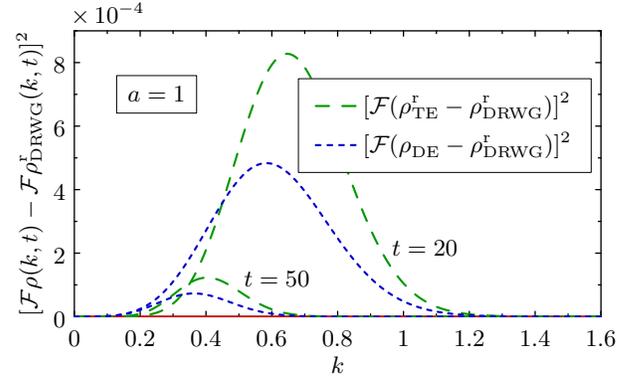}
  \caption{(Color online) The integrands of the integrals \eqref{eq:Integrals} at $t=20$ and $t=50$ with $\sigma=1$, $a=1$ and $\theta=1$, \ie, $\varepsilon=0.05$ and $\varepsilon=0.02$. Note that the vertical scale unit is $10^{-4}$.}
  \label{fig:FourHarmLongTDiscrep2}
\end{figure}

\begin{figure}[!htb]
  \centering
  \includegraphics{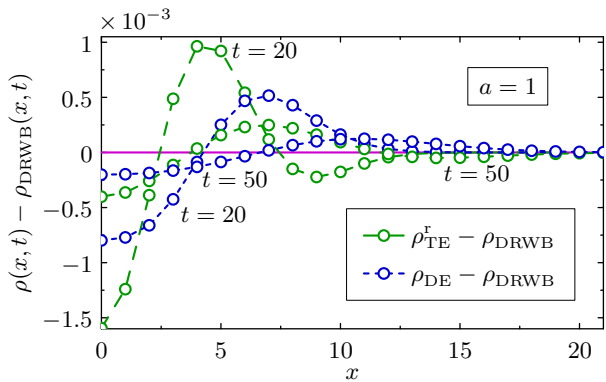}
  \caption{(Color online) The differences $\rho_\te^\regul(x_l,t) - \rho_\drwb^{}(x_l,t)$ and $\rho_\de^{}(x_l,t) - \rho_\drwb^{}(x_l,t)$ at $t=20$ and $t=50$, $\sigma=1$, $a=1$ and $\theta=1$. Note that the vertical scale unit is $10^{-3}$. Cf. with Fig.\,\ref{fig:DensDiscrepBin}.}
  \label{fig:DensDiscrepBinLongT}
\end{figure}

For comparison we show in Fig.\,\ref{fig:DensDiscrepBinLongT} differences between each
of the two approximations and the densities of the discrete-space
DCTRW (not Fourier transforms), obtained at the same long times $t=20$ and $t=50$ with $a=1$. Note that Fig.\,\ref{fig:DensDiscrepBinLongT} is qualitatively similar to Fig.\,\ref{fig:DensDiscrepBin}.
In these cases the $l_2$-discrepancy for the diffusion approximation
is also less than that for the telegraph one. We do not show
differences between the two approximations and the densities of
the continuous-space DCTRW for $t=20$ and $t=50$ because
Fig.\,\ref{fig:FourHarmLongTDiscrep2} illustrates this via the
Fourier transforms.

\begin{figure}[!htb]
  \centering
  \includegraphics{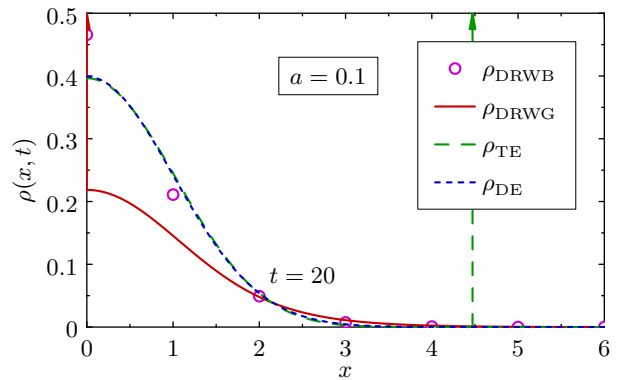}
  \caption{(Color online) The densities $\rho_\drwg^{}(x,t)$ and $\rho_\drwb^{}(x_l,t)$ and the telegraph and diffusion approximations $\rho_\te^{}(x,t)$ and $\rho_\de^{}(x,t)$ at $t=20$, $\sigma=1$, $a=0.1$ and $\theta=1$ ($\varepsilon = 0.5$). The telegraph and diffusion approximations are very close to each other. The singular terms $\rho_\drwg^\singul$ and $\rho_\te^\singul$ are depicted by vertical arrows. Note that the singular term $\rho_\drwg^\singul$ is quite ``heavy''.}
  \label{fig:DensA01SmallT}
\end{figure}

\begin{figure}[!htb]
  \centering
  \includegraphics{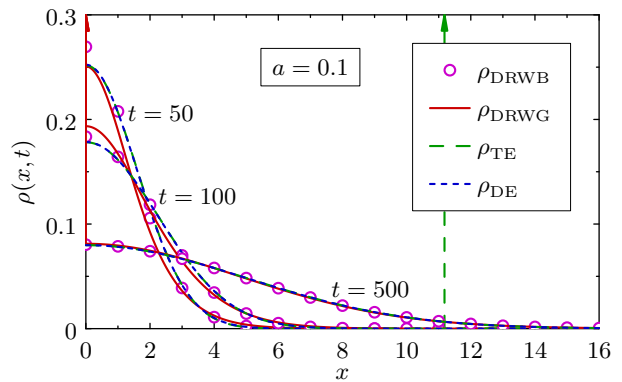}
  \caption{(Color online) The densities $\rho_\drwg^{}(x,t)$ and $\rho_\drwb^{}(x_l,t)$ and the telegraph and diffusion approximations $\rho_\te^{}(x,t)$ and $\rho_\de^{}(x,t)$ at $t=50$, $t=100$ and $t=500$, $\sigma=1$, $a=0.1$ and $\theta=1$ ($\varepsilon = 0.2$, $\varepsilon = 0.1$ and $\varepsilon = 0.02$). The telegraph and diffusion approximations are almost undistinguishable in the figure. The singular terms $\rho_\drwg^\singul$ and $\rho_\te^\singul$ (only for $t=50$) are depicted by vertical arrows. Cf. with Fig.\,\ref{fig:Dens}.}
  \label{fig:DensA01}
\end{figure}

Figs.\,\ref{fig:DensA01SmallT} and \ref{fig:DensA01} show the
densities of the continuous- and discrete-space DCTRWs and their
telegraph and diffusion approximations with the parameter
$a=0.1$. Fig.\,\ref{fig:DensA01SmallT} corresponds to the small
value of time $t=20$ ($\varepsilon = 0.5$), which is equal to the
mean waiting time. In this case the singular term
$\rho_\drwg^\singul$ is quite ``heavy'': the probability that the
particle stays at the initial point is $\varPsi_{0.1}(20) \approx
0.37 $. Note that the telegraph and diffusion
``approximations'' essentially differ from the densities of the
DCTRWs. At the same time they are very close to each other.
Fig.\,\ref{fig:DensA01} corresponds to intermediate and long
time: $t=50$, $t=100$ (intermediate) and $t=500$ (long) ($\varepsilon = 0.2$, $\varepsilon = 0.1$ and $\varepsilon = 0.02$, respectively). For $t=50$ the singular term
$\rho_\drwg^\singul$ is still ``heavy'' enough: the probability
that the particle stays at the initial point for the time $t=50$
is $\varPsi_{0.1}(50) \approx 0.08$. It is necessary to emphasize that all the results are in agreement
with the asymptotics
\eqref{eq:DTeleDCTRWFour}--\eqref{eq:DTeleDiffFour} (even for intermediate and small values of time): the
differences between the two approximations are much less than
the differences between each of them, and the densities of the
DCTRWs, while the diffusion approximation is slightly better than
the telegraph one.

The DCTRWs with small $a$ can be explained as follows. Small
values of $a$ mean that the particle mostly remains in rest than
jumps, since the
mean waiting time is long enough. Small values of time can be
described as lesser than or comparable with the mean waiting time
$2\theta/a$, \ie, such that $a t/\theta \lesssim 1$. Therefore,
for small values of time both the diffusion and telegraph
equations badly approximate the DCTRW, see
Fig.\,\ref{fig:DensA01SmallT}. At the same time, if $a t/\theta
\sim 1$ then $t/\theta \gg 1$, \ie, the singular term
\eqref{eq:DensTeleEqSingul} of the telegraph approximation is
negligible. In this case the solution of the telegraph equation
behaves like the solution of the diffusion equation. At long times
the telegraph and diffusion approximations are almost
undistinguishable and both are good, while the diffusion one is
slightly better.

\section{Conclusive remarks}
\label{sec:Concl}

We should remind that the telegraph equation describes transport
phenomena with finite propagation speed, while the velocity of the
motion of the particle in the DCTRWs is infinite, \ie, the DCTRWs
do not simulate the most distinctive property
of the telegraph equation. Moreover, asymptotic analysis and
computations performed in this paper show that the diffusion
equation gives better approximations to the DCTRWs than the
telegraph equation. In other words, \emph{the DCTRWs are always
closer to their continuous limit than to the solutions of the
telegraph equation}. This implies that, in contrast to the
widespread opinion, the DCTRWs do not simulate the telegraph equation
at the microscopic level.
Likewise, CTRWs with the exponentially distributed and general
waiting time do not simulate the telegraph equation at the
microscopic level.

Summing up, we can conclude that, first, the DCTRWs do not provide
any correct microscopic interpretation of the one-dimensional
telegraph equation, and second, the kinetic (exact) model of the
telegraph equation is different from the model based on the
DCTRWs.

An interesting question arises concerning discrete-space random
walks with the distributions of the jump length $\lambda_\discr$,
Eq.\,\eqref{eq:BernRW}, and the waiting time $\psi_\mu$,
Eq.\,\eqref{eq:PsiMu}. For sufficiently large $\mu$ this walk is
very close to the Bernoulli random walk, Eq.\,\eqref{eq:BernRW}.
For $\mu = 1$ this is the discrete-space CTRW with the
exponentially distributed waiting time $(1/\tau) \e^{t/\tau}$. In
the former case the telegraph approximation is better than the
diffusion one, while in the latter case the diffusion
approximation is better. 
Being incompletely stated yet, the problem is the following one: 
is there $\mu_0$ such that for $\mu \in
[1,\mu_0)$ the diffusion equation gives better approximation to
these discrete-space CTRWs than the telegraph one, while for $\mu
\in (\mu_0,\infty)$ the telegraph approximation is better?

\begin{acknowledgments}
The support by the RFBR grants 08-01-00315-a, 10-01-00627-a,
11-01-00573-a, the EC Collaborative Project HEALTH-F5-2010-260429 and
project no.\,14.740.11.0166 of the Russian Ministry of Science and Education 
is gratefully acknowledged.

We thank anonymous referees for useful comments.
\end{acknowledgments}

\appendix

\section[Two particular DCTRWs and the telegraph and diffusion approximations]{Two particular (discrete- and continuous-space) DCTRWs and the telegraph and diffusion approximations}
\label{sec:TwoParticul}

The density of the DCTRW with the distribution $\lambda_\Bern$ \eqref{eq:LambdaB} of
the jump length results from the discrete-space variant of the
formula \eqref{eq:DensCTRW}. We denote the density by
$\rho_\drwb$. 
It is given by
\begin{multline}
  \label{eq:DensDCTRWB}
  \rho_\drwb(x_l,t) =
  \sum_{n=|l|}^\infty p_{a,n}(t) \lambda_{\Bern,n}(x_l),\\
  x_l = l\sigma, \quad l \in \mbb{Z},
\end{multline}
where the probability density $\lambda_{\Bern,n}$ is given by the scaled binomial distribution
\begin{multline*}
  \lambda_{\Bern,n}(x_l) \\[1ex]=
  \begin{cases}
    \dfrac{n!}{2^n \,[(n+|l|)/2]! \,[(n-|l|)/2]!}, \\[2ex] \hspace{8.5em}|l| \leq n \enskip\text{and}
    &\hspace{-1ex}n+l \enskip\text{is even},\\[1.5ex]
    0, \hspace{8.2em}n < |l| \enskip\text{or}
    &\hspace{-1ex}n+l \enskip\text{is odd},
  \end{cases}
\end{multline*}
and $p_{a,n}$ stands for the probability $p_n$, corresponding to the distribution $\psi_a$. 
The Laplace transform of $p_{a,n}$ is
\begin{equation*}
  \Lapl{p}_{a,n}(s) =
  \left( \frac{a}{\theta^2} \right)^n \frac{s + 2/\theta}{[(s + 1/\theta)^2 - (1-a) / \theta^2]^{n+1}}.
\end{equation*}
Therefore, the probability $p_{a,n}$ is given by
\begin{equation*}
  p_{a,n}(t) =
  \begin{cases}
  \e^{-t/\theta} \left( \dfrac{a}{\theta^2} \right)^n \left[ f'(t) + \dfrac{1}{\theta} f(t) \right], & 0<a<1,\\[3ex]
  \e^{-t/\theta} \bigg[ \dfrac{1}{(2n)!} \left(\dfrac{t}{\theta}\right)^{2n} \\[3ex] \hspace{2.5em}+ \dfrac{1}{(2n+1)!} \left(\dfrac{t}{\theta}\right)^{2n+1} \bigg], & a=1,
  \end{cases}
\end{equation*}
where
\begin{multline*}
  f(t) =
  \frac{1}{n!} \sum_{m=0}^n \frac{(n+m)!}{m! \,(n-m)!} \,\frac{t^{n-m}}{(2 \sqrt{1-a}/\theta)^{n+m+1}}\\[1ex]
  \times \left[ (-1)^m \e^{\sqrt{1-a} \,t/\theta} + (-1)^{n+1} \e^{-\sqrt{1-a} \,t/\theta} \right]
\end{multline*}
with corresponding limit as $a \to 1$.

The density of the DCTRW with the Gaussian distribution
$\lambda_\Gauss^{}$ \eqref{eq:LambdaG} of the jump length results from the formula
\eqref{eq:DensCTRW}. We denote the density by $\rho_\drwg^{}$. 
It is given by
\begin{equation}
  \label{eq:DensDCTRWG}
  \rho_\drwg^{}(x,t) =
  \rho_\drwg^\singul(x,t) + \rho_\drwg^\regul(x,t),
\end{equation}
where the singular term is
\begin{equation}
  \label{eq:DensDCTRWGSingul}
  \rho_\drwg^\singul(x,t) = \varPsi_a(t) \delta(x),
\end{equation}
with
\begin{equation}
  \label{eq:BPsiA}
  \varPsi_a(t) =
  \begin{cases}
  \e^{-t/\theta} \bigg[ \cosh \left( \dfrac{t}{\theta} \sqrt{1-a} \right) \\[3ex] \hspace{1.5ex}+ \dfrac{1}{\sqrt{1-a}} \sinh \left( \dfrac{t}{\theta} \sqrt{1-a} \right) \bigg], & 0<a<1,\\[3ex]
  \e^{-t/\theta} \left( 1 + \dfrac{t}{\theta} \right), & a=1,
  \end{cases}
\end{equation}
and the regular term is
\begin{equation*}
  \rho_\drwg^\regul(x,t) =
  \sum_{n=1}^\infty p_{a,n}(t) \frac{1}{\sqrt{2\pi n} \,\sigma} \e^{- x^2 / (2n \sigma^2)}.
\end{equation*}

The \emph{telegraph approximation} is given by the solution to
the initial-value problem \eqref{eq:TeleEq},
\eqref{eq:TeleEqIniCond} for the telegraph equation
\cite{MorseFeshbach53v1,MoninYaglom71,Polyanin02}
\begin{equation}
  \label{eq:DensTeleEq}
  \rho_\te^{}(x,t) =
  \rho_\te^\singul(x,t) + \rho_\te^\regul(x,t),
\end{equation}
where the singular term is
\begin{equation}
  \label{eq:DensTeleEqSingul}
  \rho_\te^\singul(x,t) =
  \e^{-t/\theta} \frac{1}{2} \left[ \delta (x - v t) + \delta (x + v t) \right],
\end{equation}
and the regular term is
\begin{multline}
  \label{eq:DensTeleEqRegul}
  \rho_\te^\regul(x,t) =
  \e^{-t/\theta} \frac{H(v t - |x|)}{2 v \theta}
  \Bigg[ I_0 \left( \frac{1}{\theta} \sqrt{t^2 - \frac{x^2}{v^2}} \right)\\[1ex]
  + t \left( \sqrt{t^2 - \frac{x^2}{v^2}} \right)^{-1} \,I_1 \left( \frac{1}{\theta} \sqrt{t^2 - \frac{x^2}{v^2}} \right) \Bigg],
\end{multline}
$v = \sqrt{a} \,\sigma /(\sqrt{2} \,\theta)$ is the velocity,
$H(\cdot)$ is the Heaviside step function.

The \emph{diffusion approximation} is given by the solution to the
initial-value problem \eqref{eq:DiffEq}, \eqref{eq:DiffEqIniCond}
for the diffusion equation
\begin{equation}
  \label{eq:DensDiffEq}
  \rho_\de^{}(x,t) =
  \frac{\sqrt{\theta}}{\sqrt{\pi a t} \,\sigma} \e^{-\theta x^2 / (a \sigma^2 t)}.
\end{equation}

Note that the significant difference between the density
$\rho_\drwg^{}$ \eqref{eq:DensDCTRWG} of the DCTRW and the
solution $\rho_\te^{}$ \eqref{eq:DensTeleEq} of the telegraph
equation is that the support of $\rho_\drwg^\singul$
\eqref{eq:DensDCTRWGSingul} is localized at the starting point of
the walk ($x=0$), while the support of $\rho_\te^\singul$
\eqref{eq:DensTeleEqSingul} is localized at the moving
front ($|x|=vt$). The term $\rho_\drwg^\singul$
is negligible for $a t/\theta \gg 1$, while the term
$\rho_\te^\singul$ is negligible for $t/\theta \gg 1$, the former
condition being stronger. Note also that the support of
$\rho_\drwg^{}(\cdot,t)$ is $\mbb{R}$, while the support of
$\rho_\te^{}(\cdot,t)$ is $[-vt,vt]$. However, the latter
difference appears to be in general not so significant. Indeed,
according to the DeMoivre--Laplace and the central limit theorems
\cite{Feller68,Feller71} the solution \eqref{eq:DensDiffEq} of the
diffusion equation approximates the Bernoulli random walk
\eqref{eq:BernRW} with $h = \sqrt{a} \,\sigma$ and $\tau =
2\theta$, $t/(2\theta)$ in the expression \eqref{eq:DensDiffEq}
being the number of jumps. The solution \eqref{eq:DensDiffEq} has
the support $\mbb{R}$, while the particle involved into the random
walk has a finite velocity, nevertheless, the diffusion
approximation is a classic and widely used one to the Bernoulli
random walk.

\section{Asymptotic comparison of the telegraph and diffusion approximations to the continuous-space DCTRW at long times}
\label{sec:AsymptCompar}

In this section the long time means $a t/\theta \gg 1$, \ie, time
is longer than the mean waiting time $2\theta/a$.
We compare the telegraph and diffusion approximations to the
continuous-space DCTRW at long times by evaluation the $L_2(\mbb{R})$ norms of the differences
$\rho_\te^\regul(\cdot,t) - \rho_\drwg^\regul(\cdot,t)$ and
$\rho_\de^{}(\cdot,t) - \rho_\drwg^\regul(\cdot,t)$.
It is more convenient to consider the
$L_2(\mbb{R})$ norms of the Fourier transforms of the differences.
Recall that the $L_2(\mbb{R})$ norms of a function $f$ and its
Fourier transform $\Four{f}$ are related by $\| f
\|_{L_2(\mbb{R})}^{} = \sqrt{2\pi} \,\| \Four{f}
\|_{L_2(\mbb{R})}^{}$. Thus, we need to estimate the integrals
\begin{multline}
  \label{eq:Integrals}
  \int_0^\infty \left| \Four\rho_\te^\regul(k,t) - \Four\rho_\drwg^\regul(k,t) \right|^2 \diff{k}
  \quad\text{and}\\[1ex]
  \int_0^\infty \left| \Four\rho_\de^{}(k,t) - \Four\rho_\drwg^\regul(k,t) \right|^2 \diff{k}
\end{multline}
at long times (the integrals are taken over the interval
$(0,\infty)$, rather than $(-\infty,\infty)$, since the integrands
are even with respect to $k$).

The Fourier--Laplace transform of the density $\rho_\drwg^{}$ of
the DCTRW with the Gaussian distribution $\lambda_\Gauss^{}$ of
the jump length is the following one
\begin{multline*}
  \Four{\Lapl\rho}_\drwg^{}(k,s)\\=
  \frac{s + 2/\theta}{(s + 1/\theta)^2 - [1 - a (1 - \e^{-\sigma^2 k^2 /2})] /\theta^2},
\end{multline*}
which results directly from the Montroll--Weiss formula
\eqref{eq:MWEq} with $\Four\lambda_\Gauss$ and $\Lapl\psi_a$. This
implies the Fourier transform
\begin{multline*}
  \Four\rho_\drwg^{}(k,t) =
  \e^{-t/\theta} \left[ \cosh \bigg( \frac{t}{\theta} \sqrt{1 - a r(k)} \right)\\[1ex]
    \shoveright{+ \frac{1}{\sqrt{1 - a r(k)}} \sinh \left( \frac{t}{\theta} \sqrt{1 - a r(k)} \right) \bigg],}\\[1ex]
  r(k) = 1 - \e^{-\sigma^2 k^2 /2},
\end{multline*}
the Fourier transform of the singular term being
\begin{equation}
  \label{eq:DensDCTRWGSingulFour}
  \Four\rho_\drwg^\singul(k,t) =
  \varPsi_a(t),
\end{equation}
where $\varPsi_a$ is given by Eq.\,\eqref{eq:BPsiA}. Obviously
$\Four\rho_\drwg^\regul(k,t) \equiv \Four\rho_\drwg^{}(k,t) -
\Four\rho_\drwg^\singul(k,t) \to 0$ as $k \to \infty$.

The Fourier--Laplace transform \eqref{eq:DensTeleEqFourLapl} of
the telegraph approximation can be rewritten as
\begin{equation*}
  \Four{\Lapl\rho}_\te^{}(k,s) =
  \frac{s + 2/\theta}{(s + 1/\theta)^2 - \left( 1 - a \sigma^2 k^2 / 2 \right) /\theta^2}.
\end{equation*}
This implies the Fourier transform
\begin{multline*}
  \Four\rho_\te^{}(k,t) =
  \e^{-t/\theta} \Bigg[ \cosh \left( \frac{t}{\theta} \sqrt{1 - a \frac{\sigma^2 k^2}{2}} \right) \\[1ex]+ \left( \sqrt{1 - a \frac{\sigma^2 k^2}{2}} \right)^{-1} \sinh \left( \frac{t}{\theta} \sqrt{1 - a \frac{\sigma^2 k^2}{2}} \right) \Bigg],
\end{multline*}
where we assume that if $a \sigma^2 k^2 > 2$, then $\sqrt{1 - a
\sigma^2 k^2 / 2} = \im \sqrt{a \sigma^2 k^2 / 2 - 1}$. The
Fourier transform of the singular term is
\begin{multline}
  \label{eq:DensTeleEqSingulFour}
  \Four\rho_\te^\singul(k,t) =
  \e^{-t/\theta} \cos(v t k)\\
  \equiv \e^{-t/\theta} \cos \left( \frac{t}{\theta} \frac{\sqrt{a} \,\sigma k}{\sqrt{2}} \right).
\end{multline}
Obviously $\Four\rho_\te^\regul(k,t) \equiv \Four\rho_\te^{}(k,t) - \Four\rho_\te^\singul(k,t) \to 0$ as $k \to \infty$.

The Fourier transform of the diffusion approximation is
\begin{equation*}
  \Four\rho_\de^{}(k,t) =
  \e^{-(a t/\theta)(\sigma^2 k^2 /4)}.
\end{equation*}

The ``singular terms'' $\Four\rho_\drwg^\singul$ \eqref{eq:DensDCTRWGSingulFour} and $\Four\rho_\te^\singul$ \eqref{eq:DensTeleEqSingulFour} are exponentially small as $t \to \infty$ and, therefore, negligible on any finite interval with respect to $k$ for large $t$.

To estimate the integrals \eqref{eq:Integrals} at long times it is
convenient to introduce the small parameter $\varepsilon = (a
t/\theta)^{-1} \ll 1$ and change the variable $k$ for $\zeta
= \sigma k / (2 \sqrt{\varepsilon}) = (\sigma k /2) \sqrt{a
t/\theta}$. Then we have in the new variable the Fourier transform
of the density of the continuous-space DCTRW
\begin{multline*}
  \Four\rho_\drwg^{}(\zeta,t) =
  \e^{-1/a\varepsilon} \Bigg[ \cosh \left( \frac{\sqrt{1 - a r(\zeta)}}{a\varepsilon} \right) \\[1ex]\shoveright{+ \frac{1}{\sqrt{1 - a r(\zeta)}} \sinh \left( \frac{\sqrt{1 - a r(\zeta)}}{a\varepsilon} \right) \Bigg],}\\[1ex]
  r(\zeta) = 1 - \e^{-2 \varepsilon \zeta^2},
\end{multline*}
the ``singular term'' being $\Four\rho_\drwg^\singul(\zeta,t) =
\e^{-1/a\varepsilon} \big[ \cosh \big( \sqrt{1-a}/a\varepsilon
\big) + \sinh \big( \sqrt{1-a}/a\varepsilon \big) / \sqrt{1-a}
\,\big]$, the Fourier transform of the telegraph  approximation
\begin{multline}
  \label{eq:DensTeleEqFourZ}
  \Four\rho_\te^{}(\zeta,t) =
  \e^{-1/a\varepsilon} \Bigg[ \cosh \left( \frac{\sqrt{1 - 2 a \varepsilon \zeta^2}}{a\varepsilon} \right) \\[1ex]+ \frac{1}{\sqrt{1 - 2 a \varepsilon \zeta^2}} \sinh \left( \frac{\sqrt{1 - 2 a \varepsilon \zeta^2}}{a\varepsilon} \right) \Bigg],
\end{multline}
the ``singular term'' being $\Four\rho_\te^\singul(\zeta,t) =
\e^{-1/a\varepsilon} \cos \big( \sqrt{2/a\varepsilon} \,\zeta
\big)$, and the Fourier transform of the diffusion
 approximation
\begin{equation*}
  \Four\rho_\de^{}(\zeta,t) =
  \e^{-\zeta^2}.
\end{equation*}
The integrals \eqref{eq:Integrals}, within a factor of $2/\sigma$, become in the new variable
\begin{multline}
  \label{eq:IntegralsZ}
  \sqrt{\varepsilon} \int_0^\infty \left| \Four\rho_\te^\regul(\zeta,t) - \Four\rho_\drwg^\regul(\zeta,t) \right|^2 \diff{\zeta}
  \quad\text{and}\\[1ex]
  \sqrt{\varepsilon} \int_0^\infty \left| \Four\rho_\de^{}(\zeta,t) - \Four\rho_\drwg^\regul(\zeta,t) \right|^2 \diff{\zeta}.
\end{multline}

The ``regular term'' $\Four\rho_\drwg^\regul(\zeta,t) =
\Four\rho_\drwg^{}(\zeta,t) - \Four\rho_\drwg^\singul(\zeta,t)$
tends to zero monotonously and exponentially
as $\zeta \to \infty$, as well as the Fourier transform
$\Four\rho_\de^{}(\zeta,t)$. Therefore,
the second integral of the integrals \eqref{eq:IntegralsZ} can be
approximated with arbitrary accuracy by the integral with the
upper limit $\zeta_0$ instead of $\infty$, if $\zeta_0$ is
sufficiently large.

For small $\varepsilon$ the Fourier transform
$\Four\rho_\te^{}(\zeta,t)$ tends to zero
monotonously and exponentially on the interval
$[0,1/\sqrt{2a\varepsilon}]$, where the square root in
Eq.\,\eqref{eq:DensTeleEqFourZ} is nonnegative. The ``singular
term'' $\Four\rho_\te^\singul$ can be neglected on this interval
for small $\varepsilon$. Asymptotic behaviour of the ``regular
term'' is $\Four\rho_\te^\regul(\zeta,t) =
\Four\rho_\te^{}(\zeta,t) - \Four\rho_\te^\singul(\zeta,t) =
O(\zeta^{-1})$ as $\zeta \to \infty$.  The reason for such a
behaviour is that the regular term $\rho_\te^\regul(x,t)$
\eqref{eq:DensTeleEqRegul} is discontinuous at $|x| = vt$.
Nevertheless, due to exponential decrease of
$\Four\rho_\te^\regul(\zeta,t)$ for small $\varepsilon$ the first
integral of the integrals \eqref{eq:IntegralsZ} can be
approximated with arbitrary accuracy by the integral with the
upper limit $\zeta_0$ instead of $\infty$, if $\zeta_0$ is
sufficiently large.
Note that for any $\zeta_0$ there exists sufficiently small
$\varepsilon$ such that $\zeta_0 < 1/\sqrt{2a\varepsilon}$, \ie,
$[0,\zeta_0] \subset [0,1/\sqrt{2a\varepsilon}]$.

Thus, for small $\varepsilon$ the integrals \eqref{eq:IntegralsZ}
can be approximated with arbitrary accuracy by the integrals with
the upper limit $\zeta_0$ instead of $\infty$, where $\zeta_0$ is
sufficiently large.

The straightforward calculations imply uniform asymptotics on the
interval $[0,\zeta_0]$:
\begin{multline*}
  \Four\rho_\drwg^\regul(\zeta,t) \\[1ex]=
  \left\{ 1 + \left[ \frac{a}{2} \left( \zeta^2 - \zeta^4 \right) + \zeta^4 \right] \varepsilon \right\} \e^{-\zeta^2} + O(\varepsilon^2)\\
  \text{as}\quad \varepsilon \to 0
\end{multline*}
and
\begin{multline*}
  \Four\rho_\te^\regul(\zeta,t) =
   \left\{ 1 + \left[ \frac{a}{2} \left( \zeta^2 - \zeta^4 \right) \right] \varepsilon \right\} \e^{-\zeta^2} + O(\varepsilon^2)\\
  \text{as}\quad \varepsilon \to 0
\end{multline*}
($\Four\rho_\drwg^{}(\zeta,t)$ and $\Four\rho_\te^{}(\zeta,t)$
have the same asymptotics). Therefore, the asymptotics for the
differences are
\begin{multline}
  \label{eq:DTeleDCTRWGRegulFourZetaAsy}
  \Four\rho_\te^\regul(\zeta,t) - \Four\rho_\drwg^\regul(\zeta,t) \\[1ex]=
  - \zeta^4 \e^{-\zeta^2} \varepsilon + O(\varepsilon^2)
  \quad\text{as}\quad \varepsilon \to 0
\end{multline}
and
\begin{multline}
  \label{eq:DDiffDCTRWGRegulFourZetaAsy}
  \Four\rho_\de^{}(\zeta,t) - \Four\rho_\drwg^\regul(\zeta,t) \\[1ex]=
  - \left[ \frac{a}{2} \left( \zeta^2 - \zeta^4 \right) + \zeta^4 \right] \e^{-\zeta^2} \varepsilon + O(\varepsilon^2)
  \quad\text{as}\quad \varepsilon \to 0.
\end{multline}

Therefore, at long times (for small $\varepsilon$) the integrals \eqref{eq:IntegralsZ} are
\begin{multline}
  \label{eq:DTeleDCTRWFour}
  \sqrt{\varepsilon} \int_0^\infty \left| \Four\rho_\te^\regul(\zeta,t) - \Four\rho_\drwg^\regul(\zeta,t) \right|^2 \diff\zeta \\[1ex]\simeq
  \varepsilon^{2.5} \int_0^\infty \zeta^8 \e^{-2\zeta^2} \diff\zeta =
  \sqrt{\frac{\pi}{2}} \,\frac{7!!}{2^9} \,\varepsilon^{2.5} \\[1ex]\approx 0.257 \,\varepsilon^{2.5}
\end{multline}
and
\begin{multline}
  \label{eq:DDiffDCTRWFour}
  \sqrt{\varepsilon} \int_0^\infty \left| \Four\rho_\de^{}(\zeta,t) - \Four\rho_\drwg^\regul(\zeta,t) \right|^2 \diff\zeta \\[1ex]\simeq
  \varepsilon^{2.5} \int_0^\infty \left[ \frac{a}{2} \left( \zeta^2 - \zeta^4 \right) + \zeta^4 \right]^2 \e^{-2\zeta^2} \diff\zeta\\[1ex]
  \shoveleft{= \sqrt{\frac{\pi}{2}} \bigg[ \left( \frac{3!!}{2^7} - \frac{5!!}{2^8} + \frac{7!!}{2^{11}} \right) a^2} \\[1ex] \shoveright{+ \left( \frac{5!!}{2^7} - \frac{7!!}{2^9} \right) a + \frac{7!!}{2^9} \bigg] \varepsilon^{2.5}} \\[1ex]\approx
  (0.020 \,a^2 - 0.110 \,a + 0.257) \,\varepsilon^{2.5}.
\end{multline}
Note that at long times the same integral for the difference $\Four\rho_\te^\regul - \Four\rho_\de^{}$ is
\begin{multline}
  \label{eq:DTeleDiffFour}
  \sqrt{\varepsilon} \int_0^\infty \left| \Four\rho_\te^\regul(\zeta,t) - \Four\rho_\de^{}(\zeta,t) \right|^2 \diff\zeta \\[1ex]\simeq
  \varepsilon^{2.5} \int_0^\infty \left[ \frac{a}{2} \left( \zeta^2 - \zeta^4 \right) \right]^2 \e^{-2\zeta^2} \diff\zeta
  \approx 0.020 \,a^2 \,\varepsilon^{2.5}.
\end{multline}

The asymptotic estimates \eqref{eq:DTeleDCTRWFour} and
\eqref{eq:DDiffDCTRWFour} show that at long times the diffusion
approximation to the density $\rho_\drwg^{}$ of the DCTRW is
better in the $L_2(\mbb{R})$ norm, than the telegraph
approximation. At the same time,
for small $a$ the telegraph approximation
is almost as good as the diffusion one.

\providecommand{\noopsort}[1]{}\providecommand{\singleletter}[1]{#1}%

\end{document}